\documentstyle[epsf,seceqn]{elsart}
\begin{document}
\begin{frontmatter}
\vspace {2truein}
\title{
A rapidly converging algorithm for solving the Kohn-Sham
and related equations in electronic structure theory}

\author{ J. Auer$^\dagger$ and E. Krotscheck$^{\dagger\ddagger}$}
\address{$^{\dagger}$Institut f\"ur Theoretische Physik, Johannes Kepler
Universit\"at Linz, A-4040 Linz, Austria}
\address{$^{\ddagger}$Institute for Nuclear Theory, University of Washington,
Seattle, WA 98195-1550, USA}

\date{\today}
\begin{abstract}

We describe a rapidly converging algorithm for solving the Kohn--Sham
equations and equations of similar structure that appear frequently in
calculations of the structure of inhomogeneous electronic many--body
systems. The algorithm has its roots the Hohenberg-Kohn theorem and
solves directly for the electron density; single--particle wave
functions are only used as auxiliary quantities. The method has been
implemented for symmetric ``slabs'' of jellium as well as for
spherical jellium clusters.  Starting from very rough guesses for the
initial electron density, convergence is reached within a few
iterations. The iterations are driven by the static electric
susceptibility.

\end{abstract}
\end{frontmatter}

\section {INTRODUCTION}

Density--functional theory \cite{KohnVashishta} is
a popular method for studying properties of inhomogeneous many--electron
systems. Central to the execution of the theory is the 
solution of the Kohn-Sham equation, which
is an effective Schr\"odinger equation for a set of single--particle
wave functions $\phi_{\bf i}({\bf r})$
\begin{equation}
	-{\hbar^2\over 2m}\nabla^2\phi_{\bf i}({\bf r})
	+ V_H[\rho\,]({\bf r})\phi_{\bf i}({\bf r}) =
	\varepsilon_{\bf i}\phi_{\bf i}({\bf r})\,,
\label{eq:KohnSham}
\end{equation}
which are characterized by a set of quantum numbers ${\bf i}$.
$V_H({\bf r})$ in Eq. (\ref{eq:KohnSham}) is an effective one--body
potential that describes both the influence of an external field and
many--body interactions and depends implicitly on the density of the
system.  In density functional theory, this effective one--body
(``Hartree'')--potential is
\begin{equation}
V_H({\bf r}) = V_C({\bf r}) + V_{xc}({\bf r})
\end{equation}
where $V_C({\bf r})$ is the Coulomb potential
\begin{equation}
V_C({\bf r}) = \int d^3r' {e^2\over |{\bf r}-{\bf r}'|}
\left[\rho({\bf r}') - \rho_+({\bf r}')\right]\,,
\label{eq:VCoulomb}
\end{equation}
and $V_{xc}({\bf r})$ is the ``exchange--correlation''
correction. $\rho_+({\bf r}')$ is the charge density of the positive
background.

From the single--particle wave functions $\phi_{\bf i}({\bf r})$ one
constructs the physical one--body density
\begin{equation}
	\rho({\bf r}) = \sum_{\bf i} n({\bf i})
	\left|\phi_{\bf i}({\bf r})\right|^2\,;
\label{eq:density}
\end{equation}
the $n({\bf i})$ are the occupation numbers of the single--particle
states.

Structurally similar equations also appear in microscopic theories of
inhomogeneous, correlated electrons \cite{Surface3,Surface4}. In fact,
equations equivalent to the Kohn--Sham equation appear in the theory
of basically {\it all\/} inhomogeneous quantum many--particle systems
such as nuclei, quantum liquid clusters, and quantum liquid films. We
disregard, for the time being, non--localities like a Fock--term and
will comment on how to include such complications further below.

In this paper we will develop a rapidly converging iterative scheme
that solves, very much in the spirit of the Hohenberg-Kohn theorem,
directly for the density $\rho({\bf r})$.

It is plausible to try to solve
Eqs. (\ref{eq:KohnSham})-(\ref{eq:density}) iteratively: One
calculates, from an initial guess for the density $\rho({\bf r})$, the
single--particle orbitals $\phi_{\bf i}({\bf r})$ from
Eq. (\ref{eq:KohnSham}), from these a new single--particle density
(\ref{eq:density}), and iterates the procedure until convergence is
reached. This procedure converges well for atoms and other systems
that are dominated by an {\it external\/} field, but the convergence
can be very slow for metallic clusters, self--bound systems, or in
cases where local charge--neutrality is important.  Judicious
``mixing'' of consecutive iterations is needed
\cite{MonPe78,BrackPriv} with admixtures of the ``new'' solutions of
less than 1 percent for clusters of a few hundred electrons. A
specialized ``gradient iter\-ation method''
\cite{BLMR92,ReinhardCluster} improves the stability of the process
considerably, but still takes very many iterations to converge.  This
situation is unsatisfactory because one should think that, {\it
because\/} these systems are very stable in nature, one should be able
to mimic nature and to design a stable and rapidly converging
numerical algorithm to obtain ground state configurations.

The cause for the sometimes delicate convergence properties is rather
clear in electronic systems: The configuration with local charge
neutrality will minimize the potential energy which can, on the other
hand, be enormous if local charge neutrality is not maintained. The
quantum kinetic energy causes only a small violation of local charge
neutrality. In other words, the quantity that drives the configuration
towards the correct ground state is the potential energy, whereas the
simple iterative procedure (\ref{eq:KohnSham}), (\ref{eq:density})
focuses on the kinetic energy.

\section{Algorithm}

Our procedure to solve the coupled equation Eq. (\ref{eq:KohnSham})--%
(\ref{eq:density}) has its roots in the Hohenberg-Kohn theorem
\cite{KohnHohenberg} that states, among others, that the one--body
density $\rho({\bf r})$ is the only truly independent variable. We
therefore design an algorithm that solves, by a Newton-Raphson
procedure, directly for $\rho({\bf r})$.

Consider the functional
\begin{equation}
	F[\,\rho\,]({\bf r})\equiv
	\sum_{\bf i} n({\bf i}) \left|\phi_{\bf i}[\rho\,]({\bf r})\right|^2
	- \rho({\bf r} )
\label{eq:Deltarho}
\end{equation}
where the $\phi_{\bf i}[\rho\,]({\bf r})$ are the solutions of Eq.
(\ref{eq:KohnSham}) for the density $\rho({\bf r})$.  The exact
solution of Eq. (\ref{eq:KohnSham}) is the density that satisfies
\begin{equation}
F[\,\rho\,]({\bf r}) = 0\,.
\label{eq:Fofrho}
\end{equation}
Non-linear equations of the type (\ref{eq:Fofrho}) can be solved
iteratively by the Newton-Raphson algorithm.  Assume that we have an
$k$-th guess for the density, say $\rho^{(k)}({\bf r})$.  A next
iteration $\rho^{(k+1)}({\bf r}) \equiv \rho^{(k)}({\bf r}) +
\delta\rho^{(k)}({\bf r})$ is defined by the condition
\begin{eqnarray}
0 &=& F[\,\rho^{(k+1)}\,]({\bf r}) =
F[\,\rho^{(k)} + \delta \rho^{(k)}\,]({\bf r})
\nonumber\\
&=&  F[\,\rho^{(k)}]({\bf r}) +
\int d^3r' {\delta F\left[\,\rho\,\right]({\bf r})
\over\delta \rho({\bf r}')} \delta \rho^{(k)}({\bf r}') +
O\left((\delta\rho^{(k)})^2\right)\,.
\label{eq:Newton}
\end{eqnarray}
Defining
\begin{equation}
{\delta F\left[\,\rho\,\right]({\bf r})
\over\delta \rho({\bf r}')} \equiv -
\epsilon({\bf r},{\bf r}';0)\,,
\end{equation}
the density correction  
$\delta\rho^{(k)}({\bf r})$ is determined by
\begin{eqnarray}
F[\,\rho^{(k)}]({\bf r}) = \int d^3r'\epsilon({\bf r},{\bf r}';0)
\delta\rho^{(k)}({\bf r}')\,.
\label{eq:Inverse}
\end{eqnarray}
which is, upon discretization, a system of linear equations which can
be solved by standard methods.

The remaining task is the calculation and physical interpretation
of the kernel $\epsilon({\bf r},{\bf r}';0)$. For any set of single
particle orbitals $\{\phi_{\bf i}({\bf r})\}$ representing a
$k$-th iteration $\{\phi_{\bf i}^{(k)}({\bf r})\}$, we have
\begin{equation}
\epsilon({\bf r},{\bf r}';0)
= \delta({\bf r}-{\bf r}') - \sum_{\bf i} n({\bf i})
\left[\phi_{\bf i}({\bf r}){\delta \phi_{\bf i}^{*}({\bf r})
\over\delta\rho({\bf r}')}
+ ~c.~c.\right]\,.
\label{eq:Variation}
\end{equation}
The variations of the single--particle wave functions
are obtained by first order perturbation theory:
\begin{equation}
{\delta\phi_{\bf i}({\bf r})\over\delta\rho({\bf r}')}
 = -\int d^3 r''
\sum_{{\bf j}\,(\ne {\bf i})}{\phi_{\bf j}({\bf r})
\phi_{\bf j}^{*}({\bf r}'')\phi_{\bf i}({\bf r}'')
\over \varepsilon_{\bf j}-\varepsilon_{\bf i}}
{\delta V_H({\bf r}'')\over\delta\rho({\bf r}')}\,.
\label{eq:deltaphi}
\end{equation}
Inserting (\ref{eq:deltaphi}) in the second term of
Eq. (\ref{eq:Variation}), we find
\begin{eqnarray}
&&\sum_{\bf i} n({\bf i})\left[\phi_{\bf i}
({\bf r}){\delta\phi_{\bf i}^{*}({\bf r})\over\delta\rho({\bf r}')}
 + {\rm ~c.~c.}\right]\nonumber\\
&=& -\int d^3r''\left[\sum_{{\bf i},{\bf j}}n({\bf i})\bar n({\bf j})
{\phi_{\bf i}({\bf r})
\phi_{\bf j}^{*}({\bf r})
\phi_{\bf i}^{*}({\bf r}'')\phi_{\bf j}({\bf r}'')
\over \varepsilon_{\bf j}-\varepsilon_{\bf i}} + {\rm ~c.~c.}\right]
{\delta V_H({\bf r}'')\over\delta\rho({\bf r}')}\nonumber\\
&=&\phantom{-} \int d^3r'' \chi_0({\bf r},{\bf r}'';0)
V_{\rm p-h}({\bf r}'',{\bf r}')\,.
\label{eq:Kernel}
\end{eqnarray}
Two new quantities have been introduced in the last line of Eq.
(\ref{eq:Kernel}):
In the density variation of the Hartree--potential we recover
a local and static approximation for the effective
``particle--hole'' interaction,
\begin{equation}
 V_{\rm p-h}({\bf r},{\bf r}')\equiv 
{\delta V_H({\bf r})\over\delta\rho({\bf r}')}\,.
\label{eq:Vph}
\end{equation}
We also have identified, by inspection, the sum over states and wave
functions in Eq. (\ref{eq:Kernel}) with the zero--frequency
Lindhard--function $\chi_0({\bf r},{\bf r}';0)$ of the
non--interacting system.  The introduction of an occupation number
factor $\bar n({\bf j})\equiv 1-n({\bf j})$ above is legitimate since
all terms in the double sum where ${\bf j}$ is an occupied state
cancel out due to the antisymmetry of the energy denominator.

Hence, the Newton-Raphson procedure suggests a density correction
$\delta\rho^{(k)}({\bf r})$ given by the linear integral equation
(\ref{eq:Inverse}), where $F\left[\,\rho^{(k)}\right]({\bf r})$ is the
functional (\ref{eq:Deltarho}) and
\begin{equation}
\epsilon({\bf r}, {\bf r}';0) 
=  \delta({\bf r}-{\bf r}') - \int d^3r'' \chi_0({\bf r},{\bf r}'';0)
V_{\rm p-h}({\bf r}'',{\bf r}')
\label{eq:epsilon}
\end{equation}
is the static dielectric function of a
non-uniform electron gas \cite{PinesNoz}.

\section{Numerical Considerations}

A number of simplifications that are independent of the geometry can
be applied to the algorithm to speed up the calculation; the
justification for such simplifications is based on the observations
that

\begin{itemize}
\item The equilibration of the configuration is dominated
by the Coulomb term $V_C({\bf r})$,
\item The dielectric function $\epsilon({\bf r},{\bf r}';0)$ does
not need to be evaluated very accurately for the iterations
to converge, and
\item The ground state density is reasonably smooth.
\end{itemize}

Generally, and in particular in the local density approximation, it is
not a problem to calculate the variation $\delta V_{\rm H}({\bf
r})/\delta\rho({\bf r}')$. However, we found that the inclusion of
the exchange--correlation term  $\delta V_{\rm xc}({\bf
r})/\delta\rho({\bf r}')$ does not noticeably improve the
convergence of the procedure: It is sufficient to use the Coulomb
interaction
\begin{equation}
V_{\rm p-h}({\bf r},{\bf r}')
\approx {e^2\over\left|{\bf r}-{\bf r}'\right|}.
\label{eq:Vphapprox}
\end{equation}
This simplification allows us to use the same algorithm in microscopic
calculations \cite{Surface4} for which the computation of the
variation of the exchange--correlation energy with respect to the
density is more time consuming.  Since the Coulomb term is vastly
dominant, the non--local Fock term could also be included in the same
manner and should not cause difficulties that are worse than the
iterative solution.

Since the kernel $\epsilon({\bf r},{\bf r}';0)$ does not need to be
very accurate for the convergence of the procedure, there is also no
need to recalculate it after every iteration. This is especially true
if one is already close to the ground state density. In fact, one can
save the $L-U$ decomposition of $\epsilon({\bf r},{\bf r}';0)$, and
needs to carry out only the back-substitution step during every
iteration. Moreover, during all those iterations when $\epsilon({\bf r},{\bf
r}';0)$ is not recalculated, Eq. (\ref{eq:KohnSham}) needs to be
solved for the occupied states only.

Finally, because the ground state density should be reasonably smooth,
it is unnecessary to include very high lying single--particle states
in the state sum (\ref{eq:Kernel}). In fact, since the electron
density falls off exponentially into the vacuum, it is in most cases
sufficient to restrict the sum over particle states to those with
negative energies. Even for very crude initial densities, we found
that it is safe to include states with energies no higher than 0.2 Ry.

With these considerations taken into account, the solution of
Eq. (\ref{eq:KohnSham}) for any trial density $\rho({\bf r})$ is the
most time consuming part of the algorithm.

We conclude this section by remarking that the iterative solution
conserves the particle number due to
\begin{equation}
\int d^3 r \chi_0({\bf r}, {\bf r}';0) = 0\,.
\end{equation}
It is therefore an important precaution to start the iterations with
an initial density $\rho({\bf r})$ that has the correct particle
number.

\section{Application for spherical jellium clusters}

The feasibility of the calculation depends, of course, on the
possibility to exploit the symmetries of the ground state and to
reduce the three-dimensional eigenvalue problem (\ref{eq:KohnSham}) to
a simpler task. If this can be accomplished, the calculation of the
static dielectric function $\epsilon({\bf r},{\bf r}';0)$ and the
implementation of the Newton-Raphson iterations is equally feasible.

We have implemented the algorithm for the symmetric slabs of jellium
studied in Ref. \cite{Surface4}, and for spherically symmetric jellium
clusters which have in the past decade received much attention
\cite{Ekardt84,deHeerPhysRep,BrackPhysRep}. The convergence
properties of the procedure are basically the same; we therefore
describe the details for the cluster problem.

In this geometry, the states $\phi_{\bf i}({\bf r})$ are characterized
by a radial quantum number $i$ and the angular momentum $\ell$, {\it
i.e.\/}  ${\bf i} = \{i,\ell\}$.  We consider closed--shell clusters,
hence each state with angular momentum $\ell$ is $2(2\ell+1)$-fold
dege\-ne\-rate. Since the system is spherically symmetric, the Hartree--
(or Kohn--Sham) equations decouple
\begin{equation}
	\phi_{\bf i}\left({\bf r}\right)
	=  \phi_{i\ell} \left(r\right) Y_{\ell,m}(\theta,\phi)
	= \frac{u_{i,\ell}
	\left(r\right)}{r}Y_{\ell,m}(\theta,\phi)\,.
	\label{eq:2}
\end{equation}
Eq. (\ref{eq:KohnSham}) is a radial Schr\"odinger equation (note that
we work in atomic units)
\begin{equation}
	-{d^2 u_{i,\ell}(r)\over dr^2}
	+\frac{\ell(\ell+1)}{r^2}\,u_{i,\ell}(r)
	+V_H[\rho(r)]u_{i,\ell}(r) =\varepsilon_{i,\ell}u_{i,\ell}(r)\,
\label{eq:6}
\end{equation}
and the density is
\begin{equation}
      \rho(r)=\frac{1}{2\pi}\,\sum_{i,\ell} n(i,\ell)(2\ell+1)
	\left|\phi_{i,\ell}(r)\right|^2\,.
      \label{eq:ClusterDensity}
\end{equation}

The calculations follow exactly those of the derivation of the
previous section; all angular integrations can be carried
out due to the spherical symmetry. The angle-averaged
Coulomb potential is
\begin{equation}
V_C(r,r') =
\int {d\Omega\over 4\pi} {2\over \left|{\bf r}-{\bf r}'\right|}
= {2\over{\rm max}(r,r')}
\end{equation}
and the angle-averaged static Lindhard function is
\begin{eqnarray}
\chi_0(r,r';0) &&=\int {d\Omega\over 4\pi}\chi_0({\bf r},{\bf r}';0)
\nonumber\\
=-\frac{1}{4\pi^2}\,&&\sum_{\ell,i,j}(2\ell+1)
	n(i,\ell)\bar n(j,\ell)
	\frac{\phi_{i,\ell}(r)\, \phi_{j,\ell}(r)
	\phi_{i,\ell}(r')\, \phi_{j,\ell}(r')}
	{\varepsilon_{j\ell}-\varepsilon_{i,\ell}}\,.
\label{eq:defF}
\end{eqnarray}
In this geometry, the Newton-procedure becomes a matter of solving a
one--dimensional linear equation.

We demonstrate the working of our algorithm for the jellium model for
Na clusters with 40 and 2018 electrons. To facilitate the comparison
with earlier work, we use the local density approximation and the
energy functional of Ref. \cite{GuL76}.  The jellium background
density $\rho_+(r)$ was taken to be a step function; as a worst--case
scenario we assumed, as starting density, the {\it same\/} density
profile for the electrons.  During the iterations, the static
susceptibility was re--calculated  for the {\it first two
steps\/} only and then kept fixed.

\begin{figure}
\begin{center}
\epsfxsize=5truein
\epsffile{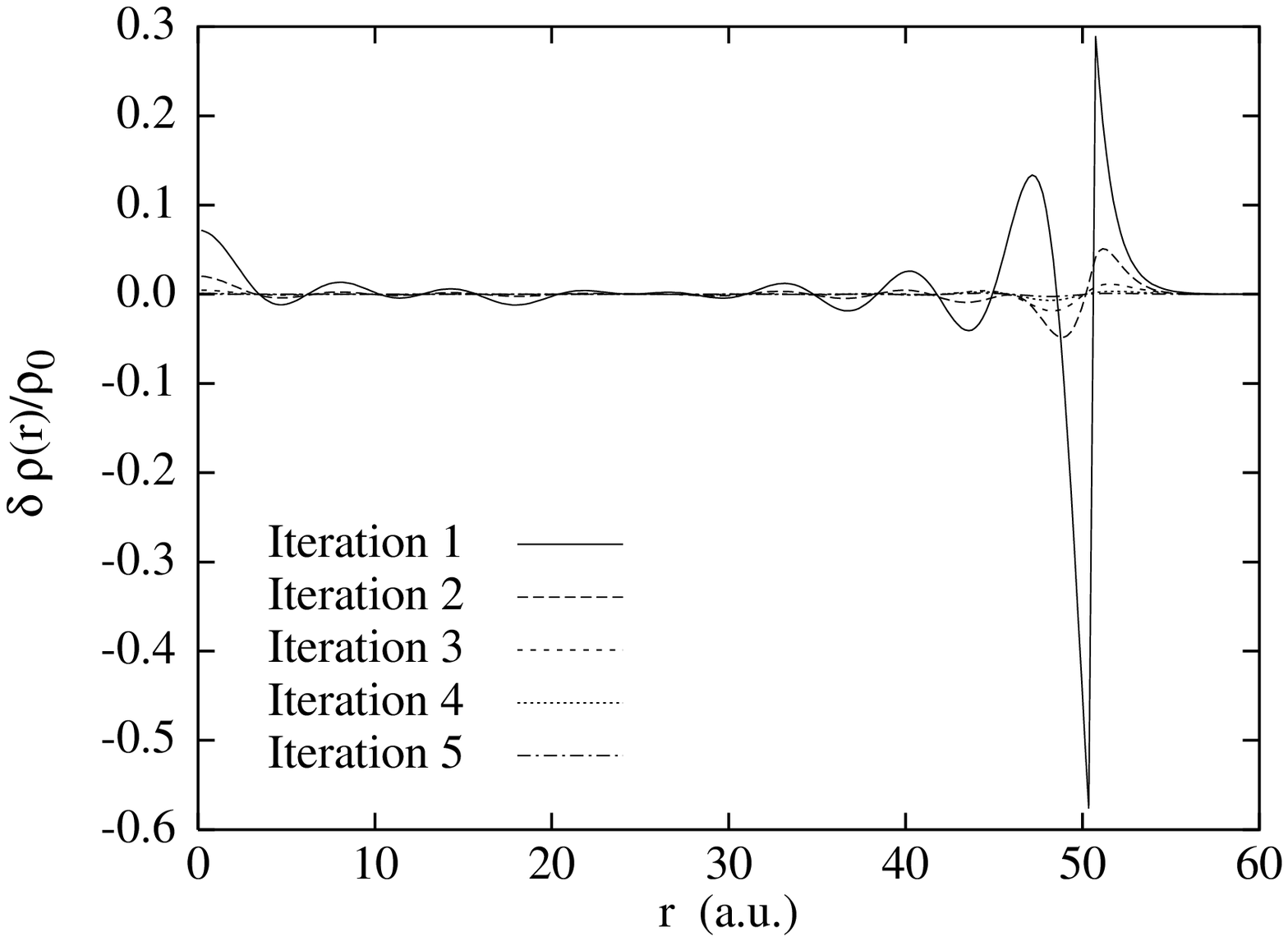}
\end{center}
\vspace{0.5truein}
\caption{The figure shows the density corrections $\delta \rho(r)/\rho_0$
for the electron density of a jellium cluster with $r_s$ = 4.0 and $N = 2018$
electrons during the first five iterations. The density is normalized
to the density $\rho_0$ of the jellium background, and the starting
density was the same as the background density. The jellium edge is
at $r = 50.55~a.u.$.}
\label{fig:iterations}
\end{figure}

\begin{figure}
\begin{center}
\epsfxsize=5truein
\epsffile{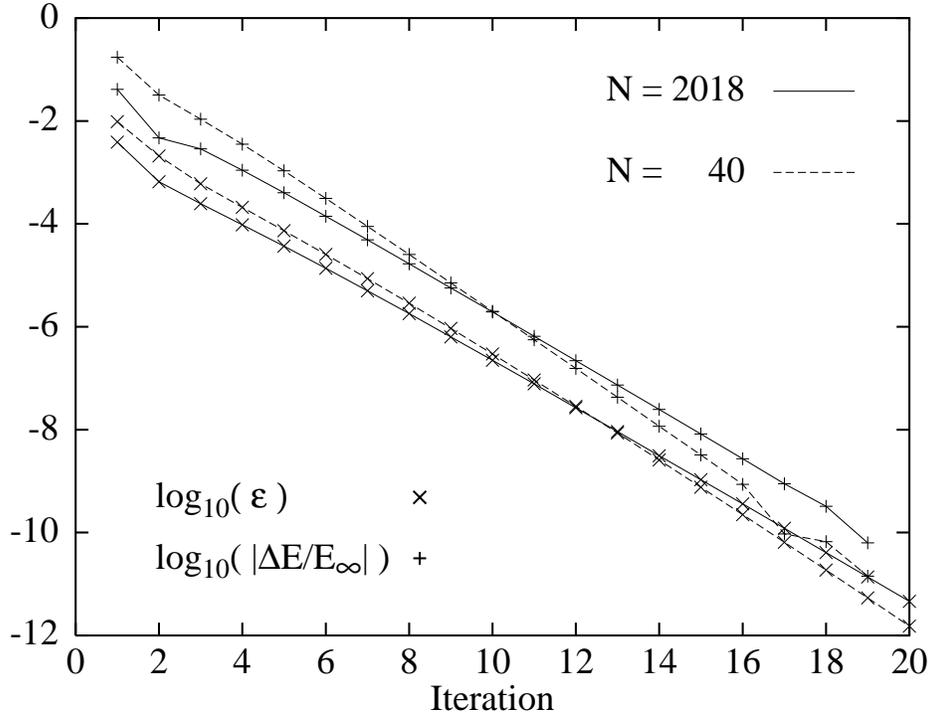}
\end{center}
\vspace{0.5truein}
\caption{The figure shows the relative energy corrections $\Delta
E/E_\infty = \left|E^{(k)}/E_\infty-1\right|$, where $E_\infty$ is the
converged result, and $E^{(k)}$ is the energy obtained in $k$-th
iteration, and the {\it rms\/} error $\epsilon =
\sqrt{\sum_i|\delta\rho(r_i)|^2}/(N\rho_0)$ of the density. Results
are shown for two jellium cluster with $r_s$ = 4.0, for $N = 2018$
(solid lines) and $N = 40$ (dashed lines).  In both cases, the
starting density was the same as the background density.}
\label{fig:convergence}
\end{figure}

Fig. \ref{fig:iterations} shows the density corrections
$\delta\rho(r)$ during the first five iterations for the 2018 electron
model; results for the average density correction as well as the
kinetic, potential, and exchange--correlation energy are given in
Table \ref{tab:iterations} as a function of the iteration number.
These results agree, within numerical accuracy, with those found in
the literature \cite{ReinhardCluster,BrackPhysRep}.  Further
documentation of the algorithm is provided in
Fig. \ref{fig:convergence} where we show the difference between the
energy obtained in the $k$-th iteration and the converged result. Also
shown is the {\it rms\/} error in the density as function of the
iteration number.  Obviously, the rate of convergence is practically
independent of the cluster size, and comparable for different
quantities of physical interest. We re-iterate that the replacement of
the Coulomb potential by the full particle--hole interaction did {\it
not\/} improve the convergence noticeably, updating the static
susceptibility after every step reduced the number of iterations
needed to obtain the accuracy shown in Fig.  \ref{fig:iterations} to
16.

\section{Summary}

We have described in this note a rapidly converging algorithm for
solving the Kohn-Sham and similar equations. While we have focused
here on electronic systems, the algorithm is not restricted to those.
Practically the same procedure has been implemented in our work on
bosonic liquid films \cite{Surface1} and droplets \cite{HNCdrops}, the
only difference being that only one state is occupied in that case. In
self--bound systems, the ``particle--hole interaction'' cannot be
replaced by the Coulomb potential, but otherwise the convergence
properties of the Newton--Raphson algorithm are the same.

For the implementation of our method, we have relied heavily on
``canned'' software, specifically BLAS and LAPACK routines for solving
the band-sym\-metric eigenvalue problems (\ref{eq:KohnSham}),
(\ref{eq:6}) and for computing the L-U decomposition of the static
dielectric function. The most time--consuming part of the algorithm is
the solution of the eigenvalue equations (\ref{eq:KohnSham}). If
needed, much computation time can be saved by choosing specialized
algorithms, such as the ``imaginary timestep method \cite{DFK80}'',
inverse iterations \cite{PFT89} or the Lanczos algorithm
\cite{KooninMeredith} that compute only the lowest lying states and
work particularly well when good estimates for these states are known.

We hasten to point out that none of the measures to speed up the
computation is essential for the implementation of our algorithm for
the simple systems considered here that can be reduced to
one--dimensional problems. Even when the static dielectric function is
updated every time, each iteration is a matter of a few seconds on a
reasonable personal computer. Efficiency considerations will become
truly useful only for higher--dimensional problems like non--spherical
clusters.

Finally, we emphasize again that a central quantity of our algorithm is
the static dielectric function $\epsilon({\bf r},{\bf r}';0)$ which is
an independently interesting physical quantity that one might wish to
compute anyway. Our algorithm also highlights the reason for the
delicate convergence of the normal iteration method: the dielectric
function is vastly dominated by the potential energy term
$\int d^3r'' \chi_0({\bf r},{\bf r}';0)
V_{\rm p-h}({\bf r}'',{\bf r}')$ whereas this term
is neglected by the conventional iterative procedure.

\begin{ack}

This work was supported, in part, by the Austrian Science Fund under
project P11098-PHY, and by the National Science Foundation under grant
No. DMR-9509743, (to S. A. Chin at Texas A\&M University).  One of us
(EK) thanks the Institute for Nuclear Theory at the University of
Washington for hospitality during the workshop on atomic clusters at
the INT in July 1998, where this research was initiated.  Discussions
and communication with G. Bertsch, M. Brack, S.-A. Chin, M. D. Miller,
S. Natschl\"ager, J. Perdew, and W. M. Saslow are gratefully
acknowledged. Special thanks go to P.-G. Reinhard for giving us access
to his code for testing purposes, and for comments on an earlier
version of this paper.

\end{ack}
\newpage
\bibliography {papers}
\bibliographystyle{prsty}
\newpage
\begin{table}
\caption{Convergence behavior of our iterations for a spherical
jellium cluster with $r_s = 4$ and $N = 2018$ electrons. All energies
are given in Rydberg units per electron. $\epsilon$ is the {\it rms\/}
error $\epsilon = \sqrt{\sum_i|\delta\rho(r_i)|^2}/(N\rho_0)$, the $r_i$ are
the mesh points, $N$ is the number of mesh points, and $\rho_0$ the
jellium density.}
\vspace{0.2truein}
\begin{center}
\begin{tabular}{cccccc}\hline\hline
Iteration &\quad$\epsilon$\quad &\quad$\left\langle T
 \right\rangle$\quad
 &\quad$\left\langle V \right\rangle$ \quad
&\quad$\left\langle E_{\rm xc}\right\rangle$\quad
& \quad $\left\langle E\right\rangle$\quad\\
\hline
  1 & 3.60E-3 &  0.14501 &  0.00000 & -0.29998 & -0.15498\\
  2 & 6.31E-4 &  0.13726 &  0.00055 & -0.29869 & -0.16087\\
  3 & 2.35E-4 &  0.13637 &  0.00065 & -0.29820 & -0.16118\\
  4 & 9.12E-5 &  0.13581 &  0.00074 & -0.29802 & -0.16147\\
  5 & 3.52E-5 &  0.13559 &  0.00078 & -0.29795 & -0.16158\\
  6 & 1.32E-5 &  0.13551 &  0.00080 & -0.29793 & -0.16162\\
  7 & 4.84E-6 &  0.13548 &  0.00080 & -0.29792 & -0.16164\\
  8 & 1.75E-6 &  0.13547 &  0.00081 & -0.29792 & -0.16164\\
  9 & 6.21E-7 &  0.13546 &  0.00081 & -0.29792 & -0.16164\\
\hline
\end{tabular}
\end{center}
\label{tab:iterations}
\end{table}
\end{document}